\documentclass[twocolumn,superscriptaddress,showpacs]{revtex4-1}
\usepackage{graphicx}
\usepackage{sidecap}
\usepackage{latexsym}
\usepackage{amssymb}
\usepackage{amsfonts}
\usepackage{soul}
\usepackage{color}
\usepackage{amsmath}

\begin{document}

\title{Tunable artificial vortex ice in nanostructured superconductors with\\ frustrated kagome lattice of paired antidots}

\author{C. Xue}
\affiliation{School of Mechanics, Civil Engineering and Architecture, Northwestern Polytechnical University, Xi'an 710072, China}
\affiliation{INPAC--Institute for Nanoscale Physics and Chemistry, KU Leuven, Celestijnenlaan 200D, B--3001 Leuven, Belgium}
\author{J.-Y. Ge}
\affiliation{INPAC--Institute for Nanoscale Physics and Chemistry, KU Leuven, Celestijnenlaan 200D, B--3001 Leuven, Belgium}
\affiliation{Materials Genome Institute, Shanghai University, Shangda Road 99, 200444 Shanghai, China}
\author{A. He}
\affiliation{College of science, Chang'an University, Xi'an 710064, China}
\author{V. S. Zharinov}
\affiliation{INPAC--Institute for Nanoscale Physics and Chemistry, KU Leuven, Celestijnenlaan 200D, B--3001 Leuven, Belgium}
\author{V. V. Moshchalkov}
\affiliation{INPAC--Institute for Nanoscale Physics and Chemistry, KU Leuven, Celestijnenlaan 200D, B--3001 Leuven, Belgium}
\author{Y. H. Zhou}
\affiliation{Key Laboratory of Mechanics on Disaster and Environment in Western China attached to the Ministry of Education of China,\\and Department of Mechanics and Engineering Sciences, Lanzhou University, Lanzhou 730000, China}
\affiliation{School of Aeronautics, Northwestern Polytechnical University, Xi'an 710072, P. R. China}

\author{\\A. V. Silhanek}
\affiliation{Experimental Physics of Nanostructured Materials, Q-MAT, CESAM, Universit\'{e} de Li\`{e}ge, B-4000 Sart Tilman, Belgium.}
\author{J. Van de Vondel}
\affiliation{INPAC--Institute for Nanoscale Physics and Chemistry, KU Leuven, Celestijnenlaan 200D, B--3001 Leuven, Belgium}
\date{\today}

\begin{abstract}
Theoretical proposals for spin ice analogs based on nanostructured superconductors have suggested larger flexibility for probing the effects of fluctuations and disorder than in the magnetic systems. In this work, we unveil the particularities of a vortex ice system by direct observation of the vortex distribution in a kagome lattice of paired antidots using scanning Hall probe microscopy. The theoretically suggested vortex ice distribution, lacking long range order, is observed at half matching field ($H_{1}/2$). Moreover, the vortex ice state formed by the pinned vortices is still preserved at $2H_{1}/3$. This unexpected result is attributed to the introduction of interstitial vortices at these magnetic field values. Although the interstitial vortices increase the number of possible vortex configurations, it is clearly shown that the vortex ice state observed at $2H_{1}/3$ is less prone to defects than at $H_{1}/2$. In addition, the non-monotonic variations of the vortex ice quality on the lattice spacing indicates that a highly ordered vortex ice state cannot be attained by simply reducing the lattice spacing. The optimal design to observe defect free vortex ice is discussed based on the experimental statistics. The direct observations of a tunable vortex ice state provides new opportunities to explore the order-disorder transition in artificial ice systems.
\end{abstract}

\maketitle

The interplay of competing forces in an ensemble of repulsive `particles' on a potential-energy landscape is ubiquitous in many physical systems. Whenever there is an impossibility to minimize all pairwise interaction, frustration emerges, which is a well-known source of degeneracy, disorder, and inhomogeneities. Frustration is the main responsible mechanism giving rise to glasses, characterized by structural disorder, and ices where the structural order is retained at expenses of a subtle balance between competing interactions. In the latter case, the limited choices to allocate pairwise interactions manifest themselves in ice-rules and give rise to a multiplicity of ground states resulting in a finite macroscopy entropy at the lowest accessible temperatures \cite{Spinice_Nisoli2}.

During the last decade, lithographically defined magnetic systems have been introduced to explore the physics of frustrated systems \cite{Spinice_wang, Spinice_moller, Spinice_Chern, Spinice_Lammert, Spinice_Rougemaille, Spinice_Nisoli1, Spinice_Nisoli2, Spinice_Zhang1, Spinice_Zhang2, Spinice_Gilbert1, Spinice_Gilbert2, Spinice_Ladak, Spinice_Branford, Spinice_Morgan, Spinice_Farhan1, Spinice_Farhan2, Spinice_Marrows, Spinice_Heyderman, Spinice_Budrikis, Spinice_Kapaklis, Spinice_Wang, Spinice_Bhat, Spinice_Qi, Spinice_Phatak, Spinice_Arnalds, Spinice_Wysin, Spinice_Porro, Spinice_Daunheimer, Spinice_Anghinolfi}. The advantages of these tailor-made systems are two-fold. On the one hand, they allow a large tunability of the system parameters (magnetic moment, array periodicity and symmetry, geometrical shape, etc). On the other hand, the fabricated single-domain ferromagnetic structures mimicking an artificial giant Ising spin can be directly visualized, thus permitting one to count the individual microscopic configurations and directly access the statistics of the ensemble.

Besides water ice \cite{Water_ice} and spin ice systems \cite{Spinice_wang, Spinice_moller, Spinice_Chern, Spinice_Nisoli1, Spinice_Zhang1}, it has been recognized that analogous ice states can exist in other systems, such as colloidal artificial ice \cite{Colloidal_ice1,Colloidal_ice1.5,Colloidal_ice2,Colloidal_ice3,Colloidal_ice4,Colloidal_ice5} and Coulombic charge ice \cite{Charge_ice}. More recently, Lib\'{a}l \emph{et al}. \cite{vortexice_libal} proposed and investigated theoretically artificial vortex ice states in a nanostructured superconductor with square and kagome lattice consisting of double-well pinning sites. The numerical simulations show that the strong repulsive vortex-vortex (V-V) interactions can drive the vortex system into the ground state more readily than in the magnetic systems. Furthermore, the tunability of these systems exceeds by far that of the magnetic counterparts as the number of vortices and vacancies can be adjusted by merely changing the external field. By performing transport measurements, the square vortex ice has been indirectly confirmed \cite{vortexice_Latimer} and it has been found that the vortex system provides interesting opportunity to freeze and thaw artificial ice by switching on/off geometric frustration via temperature changes \cite{vortexice_Trastoy}. By using scanning Hall probe microscopy (SHPM), it was found that the filling rules of degenerate vortex configurations in a kagome lattice of elongated antidots are reminiscent of the ice rules \cite{vortex_degeneracy}. Very recently, the square vortex ice state has been visualized using SHPM \cite{vortexice_Ge}, confirming the possibility of this system to create defects by tuning the magnetic field. Despite the progress achieved in identifying and imaging the vortex ice states in a square lattice, the direct visualization and stability analysis of a vortex ice as a function of the lattice parameters has not been addressed yet.

In the present work, using SHPM, we directly probe the formation and stability of the vortex ice state as a function of the applied magnetic field by performing consecutive field-cooling (FC) experiments in four samples with different kagome lattice parameters. Besides confirming the theoretical predictions, we unveil novel features, unique to the vortex ice system: (i) The vortex ice state, formed by the pinned vortices, persists at $2H_{1}/3$ due to the extra degree of freedom induced by the interstitial vortices. The obtained vortex ice state is more robust against the formation of defects, such as empty-pairs and saturated-pairs. (ii) The obtained statistics on different samples, regarding the presence of defects, clearly indicates that the ordered vortex ice states cannot be attained by simply reducing the lattice spacing. Using these insights, we further discuss the optimal design required to obtain a highly ordered vortex ice state.

\begin{figure}[!t]
\centering
\includegraphics*[width=0.8\linewidth,angle=0]{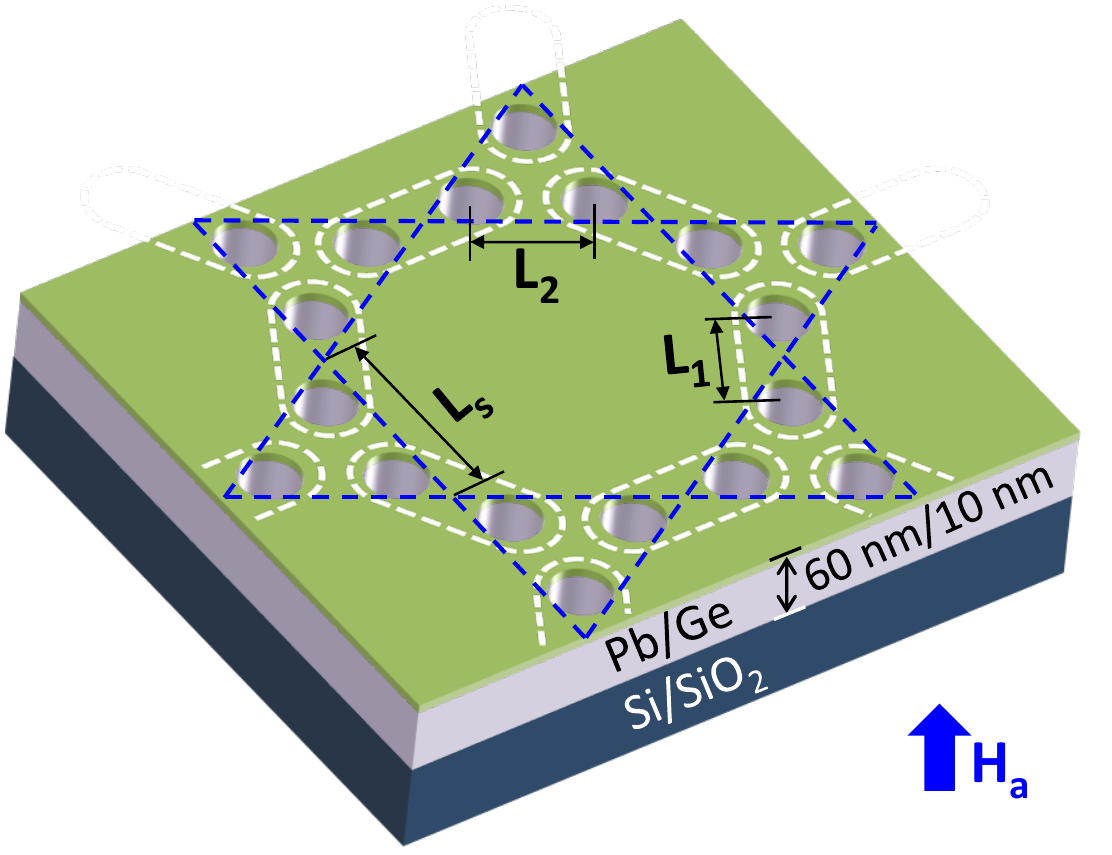}
\caption{Schematic representation of the nanostructured superconducting sample (top green layer) patterned with a kagome lattice consisting of paired antidots (white dashed rounded ellipses) on Si/SiO$_{2}$ substrate (bottom blue layer). The paired vortex-vortex (V-V) distance and the nearest neighbour V-V distance at each vertex of the kagome lattice are denoted by $L_{1}$ and $L_{2}$, respectively. The magnetic field $H_{a}$ is perpendicular to the samples.}
\end{figure}

The investigated nanostructured kagome lattices are prepared using conventional electron-beam lithography. As shown in Fig. 1, the kagome lattice with double-well sites consists of adjacent pairs of antidots with a center-to-center distance of $L_{1}$ (marked by white ellipses). The paired antidots are placed on the side of a hexagon and three antidots, with a center-to-center distance of $L_{2}$, meet at the vertex of the lattice. The resulting kagome lattice spacing is $L_{s}$ =$\sqrt{3}L_{1}/2+L_{2}$. As a consequence, the vortex configurations and the occupation number at each vertex are mainly determined by two types of nearest neighbour interactions. Bearing in mind the resolution and scan size of the SHPM we designed four different variations of the kagome antidot lattice in a Pb film with the same nominal thickness of 60 nm and a 10 nm Ge layer on top to prevent oxidation. Subsequently, the samples are covered by a layer of 35 nm-thick Au as a conducting layer for allowing the approaching of the Hall sensor via a scanning tunneling probe. The exact dimensions of these four samples and their first matching field $H_{1}$ are indicated in Ref. \cite{Supple}. All presented SHPM images are obtained by retracting the scanning tunneling probe 400 nm after approaching the sample surface at $T$ = 4.25 K.

In a first step, we will unveil the ingredients responsible for the formation of vortex ice states and its defects, by exploring the vortex distribution in sample-III with $L_{1}=L_{2}=2$ $\mu m$ (see Fig. 2(a)). Fig. 2(b) shows the vortex distributions observed at different applied magnetic fields ($H_{a}=H_{1}/3$, $H_{a}=5H_{1}/6$ and $H_{a}=1.53H_{1}$). At $H_{a}=1.53H_{1}$ (last panel of Fig. 2(b)) all antidots are occupied by vortices, while the interstitial vortices are constrained in a caging potential produced by the pinned vortices. As such, the exact position of antidots in the scanned area can be determined based on these vortex distributions \cite{Supple}. At $H_{a}=H_{1}/3$, only one antidot at each vertex is occupied (one-occupied/two-empty), i.e. $n_{in} = 1$  ($n_{in}$ is defined as the number of `in' vortices at a vertex). In addition, paired antidots are never simultaneously occupied due to the high energy associated to this configuration. At $5H_{1}/6$, an interstitial vortex is observed in each hexagon and two thirds of the antidots are occupied by vortices. As a result, the vortices pinned in the antidots comply with two-occupied/one-empty configuration at each vertex ($n_{in} = 2$). Similar with our recent experiments in kagome lattice of elongated antidots \cite{vortex_degeneracy}, the vortex arrangements with $n_{in} = 1$ or $n_{in} = 2$ can leads to degeneracy and a large configuration entropy. It is worth noting that although the ice state is defined by the pinned vortices, interstitial vortices are already present before half matching field. The retaining zero-point entropy in the stuffing of Ho$_{2}$Ti$_{2}$O$_7$ with extra Ho ions suggests that the ice rules have relevance beyond the pure spin-ice system \cite{stuffed_ice}. In the kagome lattice, the amount of pinned vortices is not linearly increasing with the applied magnetic field and therefore the impact of the interstitial vortices on the vortex ice state has to be taken into account.

In order to explore the impact of the aforementioned effects on the observed vortex ice states, we perform FC measurements at four different locations of Sample-III at $H_{1}/2$, which are shown in the upper panels of Fig. 2(c). Despite the lack of an overall ordered vortex ground state, some common topological characteristics can be identified. In order to clarify the observed patterns, a schematic representation of the vortex distribution is given in the lower panels for each location (black, white and red dots represent a pinned vortex, a vacancy and an interstitial vortex, respectively). As seen from this representation each vertex is occupied by either one ($n_{in} = 1$) or two vortices ($n_{in} = 2$), analogous to the ice states in spin ice systems \cite{Spinice_Nisoli2,Spinice_Chern,Spinice_Rougemaille,Spinice_Qi,Spinice_Arnalds,Spinice_Daunheimer}. The absence of $n_{in} = 0$ and $n_{in} = 3$ defects in our measurements is in perfect agreement with numerical simulations for a kagome vortex ice system \cite{vortexice_libal}. Nevertheless, due to the presence of interstitial vortices at $H_{1}/2$, the number of pinned vortices is lower than half of the number of antidots and, as a result, some of the paired antidots are still completely empty. This is a type of defect unique for vortex systems \cite{vortexice_Ge} and has  no correspondence in a spin ice system.

\begin{figure}[!t]
\centering
\includegraphics*[width=1.0\linewidth,angle=0]{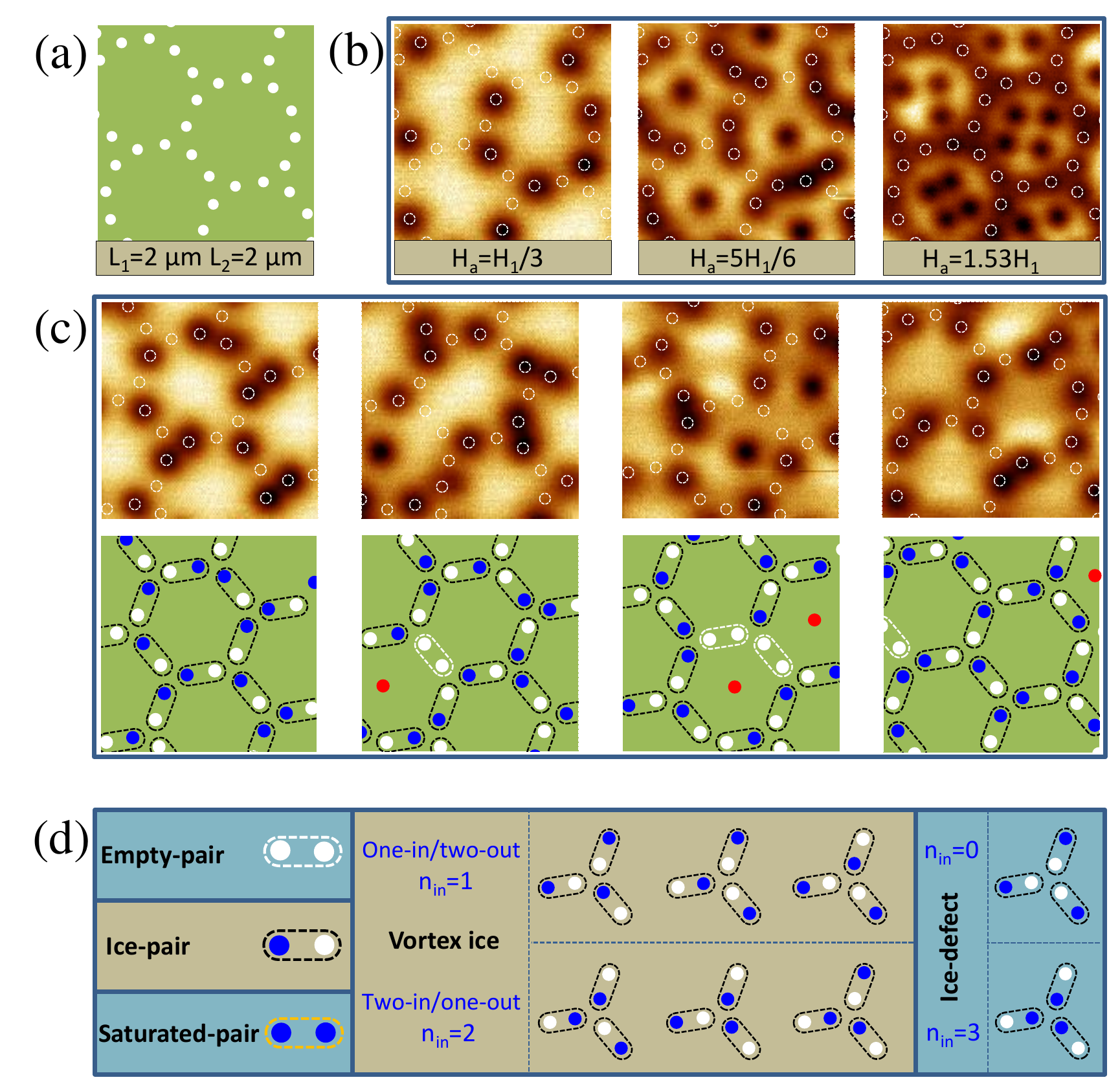}
\caption{\textbf{(a)} Optical image (16 $\times$ 16 $\mu m^{2}$) of Sample-III with lattice parameters: $L_{1}$ = 2 $\mu m$ and $L_{2}$ = 2 $\mu m$. \textbf{(b)} Vortex configurations in Sample-III at $H_{a}$ = $H_{1}/3$, $5H_{1}/6$, and $1.53H_{1}$. The positions of the antidots are marked by white circles. \textbf{(c)} The observed vortex ice states (upper panels) and their schematic representation (bottom panels) at four different locations of Sample-III for $H_{a}=H_{1}/2$, where the vortices obey $n_{in}$=1 or $n_{in}$=2 at each vertex. The interstitial vortices are marked by red circles. \textbf{(d)} Sketch of vortex-occupations in paired antidots: empty-pair (no vortex in paired antidots, surrounded by white dashed ellipse), ice-pair (only one vortex in paired antidots, black dashed ellipse), and saturated-pair (two vortices in paired antidots, orange dashed ellipse). Six possible vortex ice configurations for each vertex, which is analogous to the ice rules, i.e. one-in/two-out (upper) and two-in/one-out (bottom). The ice-defects with $n_{in}$ = 0 and $n_{in}$ = 3 (rightmost sketches).}
\end{figure}

\begin{figure}[!t]
\centering
\includegraphics*[width=0.94\linewidth,angle=0]{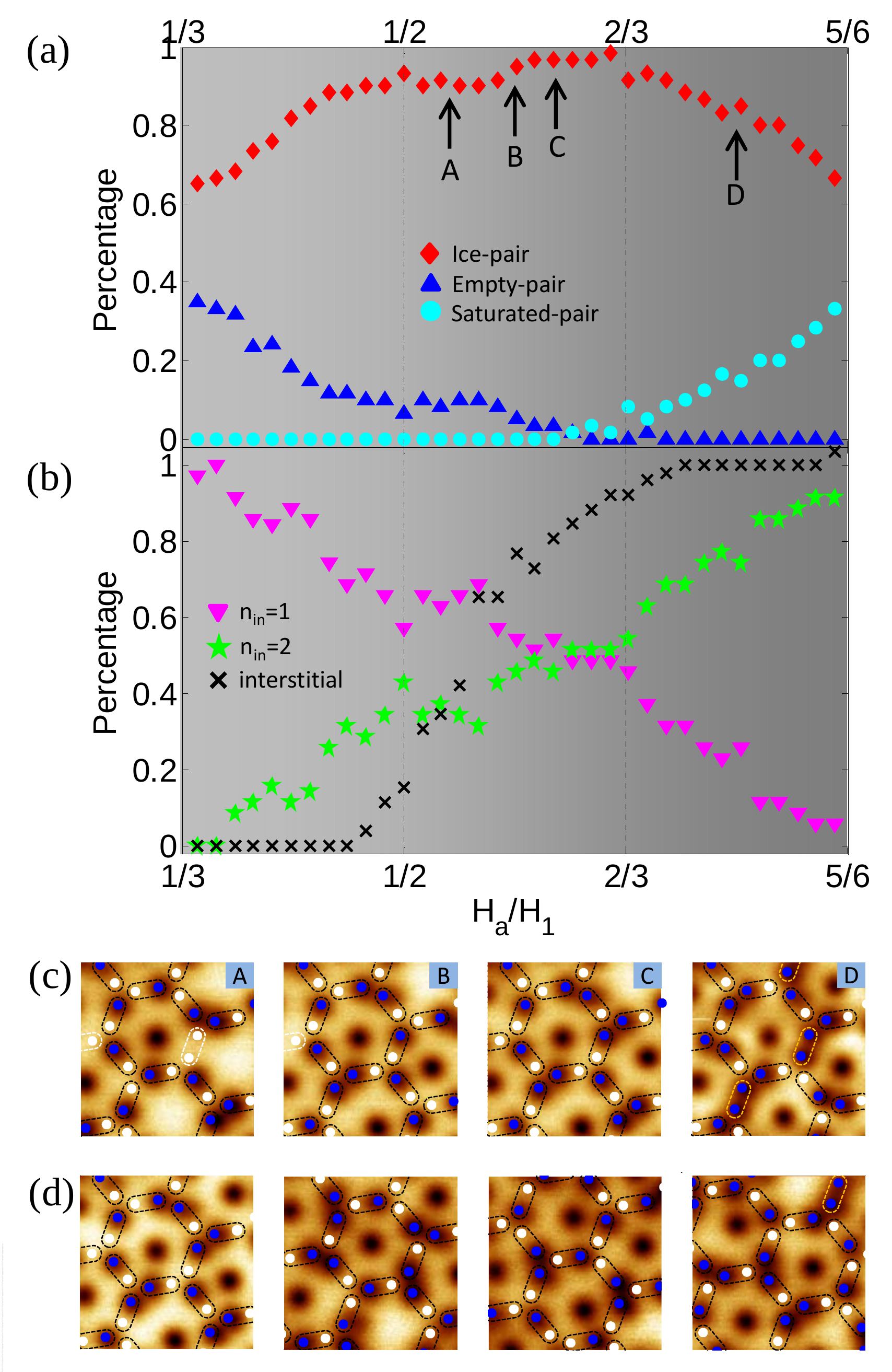}
\caption{\textbf{(a)} The occurrence (in percentage) of ice-pairs (red diamond), empty-pairs (blue triangle), and saturated-pairs (cyan circle) vs external magnetic field based on the statistics of vortex-occupations in paired antidots (See Ref. \cite{Supple}: original SHPM images of vortex states at different locations of Sample-III under various magnetic fields). \textbf{(b)} The occurrence (in percentage) of the different occupation numbers for each vertex as a function of the applied magnetic field. Purple triangle: $n_{in}$=1; green star: $n_{in}$=2; black cross: interstitial vortex. \textbf{(c)} Vortex configurations obtained after field-cooling at different magnetic fields as indicated by the capital letters in (a). \textbf{(d)} The vortex ice states with additional interstitial vortices observed at $H_{a}$ = 2$H_{1}/3$ in four different locations of Sample-III (upper panels).}
\end{figure}

\begin{figure*}[!t]
\centering
\includegraphics*[width=1.0\linewidth,angle=0]{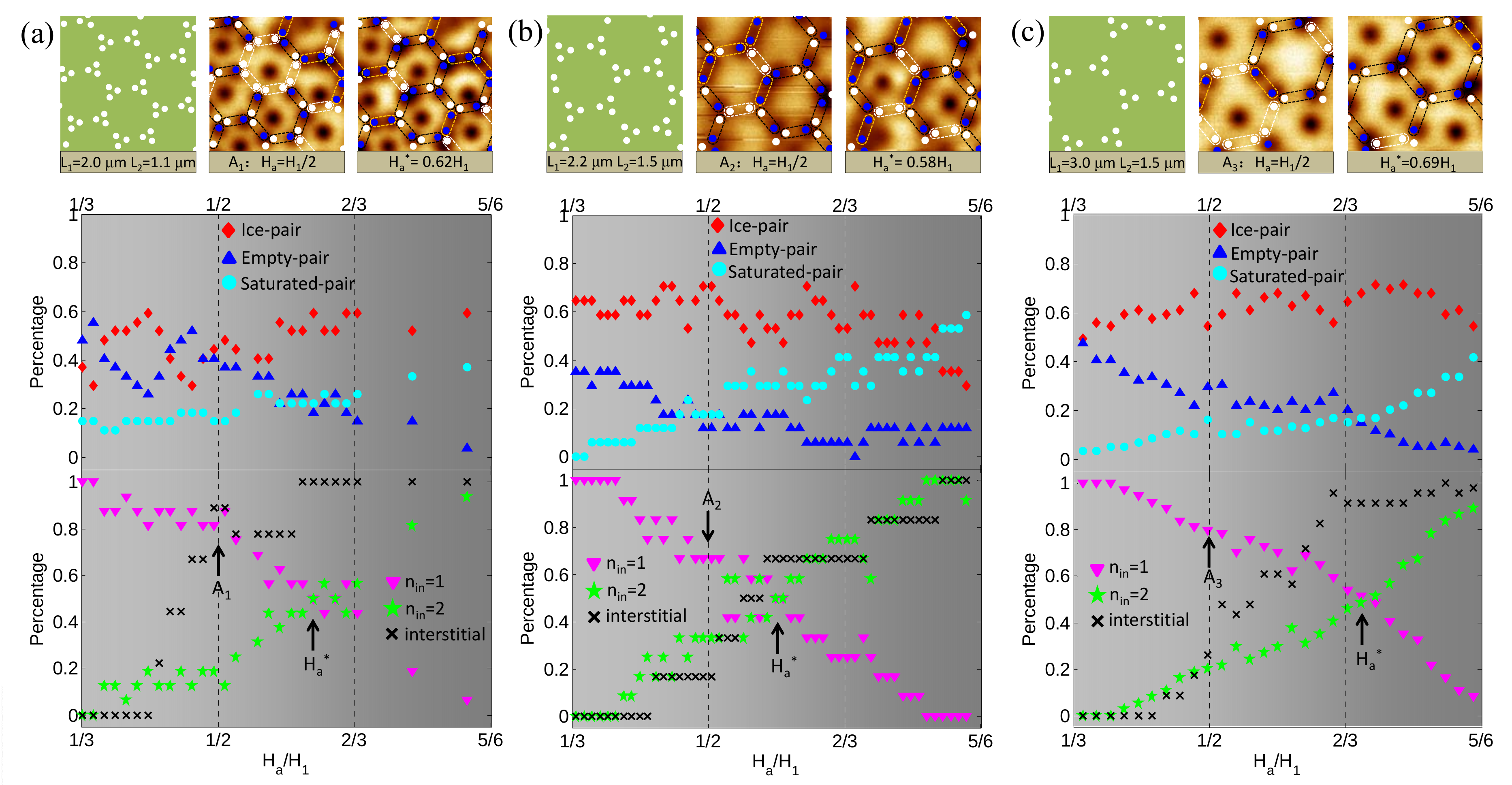}
\caption{Statistics of vortex ice states in Sample-I with $L_{1}$ = 2 $\mu m$, $L_{2}$ = 1.1 $\mu m$ \textbf{(a)}, Sample-II with $L_{1}$ = 2.2 $\mu m$, $L_{2}$ = 1.5 $\mu m$ \textbf{(b)}, and Sample-IV with $L_{1}$ = 3 $\mu m$, $L_{2}$ = 1.5 $\mu m$ \textbf{(c)}. Upper left: schematic representations of the sample with same size as scanned area (16 $\times$ 16 $\mu m^{2}$). Upper middle: The vortex configurations observed in three samples at $H_{a}$ = $H_{1}/2$ (points $A_{1}$, $A_{2}$ and $A_{3}$ in bottom panels). Upper right: The vortex ice states at a magnetic field value $H_{a}^{*}$ defined by the constrained: $n_{in}=1 \approx n_{in}=2$  (points $H_{a}^{*}$ in bottom panels). Bottom panels: The occurrence (in percentage) of the different pair distributions (ice-pairs (red diamond), empty-pairs (blue triangle) and saturated-pairs (cyan circle)) and the different vertex distributions ($n_{in}$=1 (purple triangle) and $n_{in}$=2 (green stars)).  The black crosses mark averaged number of interstitial vortices per hexagon.}
\end{figure*}

Before we continue this discussion, we first introduce some important terms needed to describe the peculiarities of a vortex ice system. As shown in Fig. 2(d), the ice-pair exhibits two possible states and can therefore be mapped into a spin system. However, there are two more possible vortex-occupations in paired antidots, namely empty-pairs without any trapped vortices and saturated-pairs with double occupation which cannot be mapped onto the spin ice systems. As a result, the generalized definition of vortex ice state in a kagome lattice is that vortex-occupations comply with $n_{in} = 1$ or $n_{in} = 2$ at all vertices \cite{vortexice_libal}. Fig. 2(d) shows six possible unit cells for vortex ice states and two possible ice-defects. Besides the percentage of vertices with $n_{in} = 1$ and $n_{in} = 2$, the percentage of ice-pairs, empty-pairs and saturated-pairs are also important physical parameters to characterize the quality of the vortex ice systems.

The kagome vortex ice system is less rigid than the magnetic systems and offers a flexible playground to change the amount of defects by changing the applied magnetic field \cite{vortexice_Ge}.  In addition, the presence of interstitial vortices even at magnetic field values below $H_1$ introduces a non-trivial relationship between the applied magnetic field and the number of pinned vortices. In the next step we explore the non-trivial dependence of the different types of defects and the quality of the vortex ice on the applied magnetic field. This is done by performing consecutive measurements at various magnetic fields and at different places of Sample-III (Fig. 3). In order to quantify the evolution of the vortex distribution with magnetic field, we extracted the occurrence (in percentage) of each possible state for the paired antidots (Fig. 3(a)) and the vertices (Fig. 3(b)). The original data, used to extract this information, are shown in Ref. \cite{Supple}. At $H_{1}/2$ nearly 10\% of empty-pair defects are observed (Fig. 3(a), blue triangles), which was attributed to the presence of interstitial vortices. Fig. 3(c) represents the vortex configurations observed at different magnetic field values exceeding $H_1/2$ (marked by the points A, B, C, and D in Fig. 3(a)). One can see that the number of ice-pairs and vertices with $n_{in} = 2$ (empty-pairs and vertices with $n_{in} = 1$) increases (decreases) almost linearly with magnetic field.  This indicates that more and more empty-pairs are occupied by vortices and become ice-pairs in the range of $H_{1}/2 < H_{a} < 2H_{1}/3$ (see the panels A, B, and C). Although the vortex states comply with $n_{in} = 1$ or $n_{in} = 2$ at all vertices for $H_{a}=H_{1}/2$ (see Fig. 2(c)), the presence of empty-pairs leads to a percentage of ice-pairs below 100\% and an imbalance between $n_{in} = 1$ and $n_{in} = 2$ (i.e. $P_{n_{in} = 1} > P_{n_{in} = 2}$). Because the distance between interstitial and pinned vortices is longer than the nearest neighbour distance between pinned vortices, the amount of interstitial vortices increases rapidly with $H_{a}$. The ice-pairs (empty-pairs) continue to increase (decrease) until $P_{ice-pair} \approx$ 1 ($P_{empty-pair} = 0$) at $2H_{1}/3$. Additionally, the occurrence (in percentages) of vertices with $n_{in} = 1$ and $n_{in} = 2$ are equal (i.e. 50\%). A set of vortex distributions, with an additional intersticial vortex in the center of the hexagon, can be observed in four different locations of sample-III at $H_{a}=2H_{1}/3$ (see Fig. 3(d)). The vortex configurations of the pinned vortices comply with the ice rules, i.e. $n_{in} = 1$ or $n_{in} = 2$ at each vertex. The similar vortex arrangements at vertex were also observed in a kagome lattice of elongated antidots \cite{vortex_degeneracy}. The percentage of vertices with $n_{in} = 1$ and $n_{in} = 2$ are nearly the same and, as stated before, empty-pair and saturated-pair defects are rarely observed ($<1.7\%$). Therefore, we can clearly state that the nearly perfect vortex ice state is facilitated by the pinned vortices, which is more robust against defects than the vortex ice states at $H_{1}/2$. This behavior can be summarized as follows. On the one hand, increasing the applied magnetic field reduces the presence of empty-pairs and increases the amount of vertices with $n_{in}=2$. On the other hand, saturated-pairs are still avoided and excessive vortices will be pushed into the interstitial positions.

\begin{figure}[!t]
\centering
\includegraphics*[width=0.85\linewidth,angle=0]{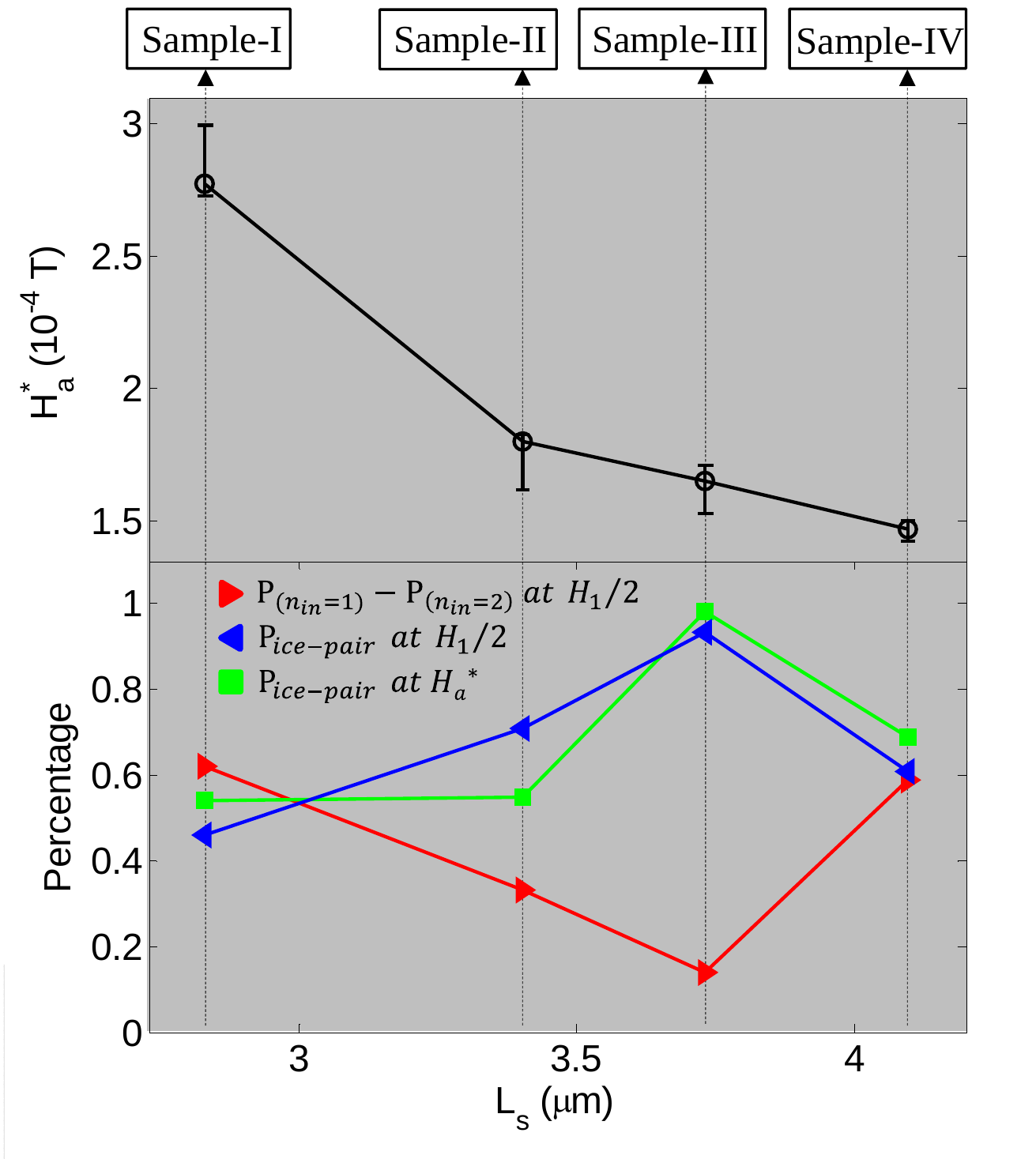}
\caption{The variation of vortex ice with the kagome lattice spacing $L_{s}$. Black circle: the magnetic field value ($H_{a}^{*}$) at which $P_{(n_{in}=1)} = P_{(n_{in}=2)}$; red triangle: the difference between the percentage of $n_{in}=1$ and $n_{in}=2$ at $H_{1}/2$; blue triangle and green square: percentage of ice-pairs at $H_{1}/2$ and $H_{a}^{*}$ respectively.}
\end{figure}

It is well-known that the V-V interaction, and consequently the resulting vortex distribution, in a superconductor strongly depends on the pinning landscape, the temperature, etc. In the kagome lattice, the V-V interactions can be tuned by changing $L_{1}$ and/or $L_{2}$. In order to explore the evolution of vortex ice states by tuning the kagome lattice parameters we further perform FC measurements on different samples with different lattice parameters. The observed vortex states are shown in Ref. \cite{Supple}. Fig. 4 shows the vortex ice states at certain magnetic fields and the occurrence (in percentage) for the different states of the antidot pairs and the vertices as a function of the magnetic field. By analyzing and comparing the obtained experimental results in Figs. 3 (Sample-III) and 4 (Samples-I, II, and IV), one can identify some common characteristics for all samples. (i) The vortex distribution is identical for all samples at $H_{1}/3$ (one-in at each vertex) and $5H_{1}/6$ (two-in at each vertex). (ii) $P_{n_{in}=1} + P_{n_{in}=2} = 100\%$ or, in other words, the vertex defects ( i.e. $n_{in} = 0$ and $n_{in} = 3$) are not observed in the magnetic field range $H_{1}/3 < H_{a} < 5H_{1}/6$ although the distance between pinned vortices is quite large in sample-IV. This indicates that the V-V interaction in the vortex ice system is much stronger than the interactions of magnetic bars in spin ice systems \cite{vortexice_libal,vortexice_Latimer,vortexice_Trastoy}.

The impact of $L_2$ can be seen by comparing sample-I ($L_{1}$=2 $\mu m$, $L_{2}$=1.1 $\mu m$) with sample-III ($L_{1}$=2 $\mu m$, $L_{2}$=2 $\mu m$). Although the interaction between vortices pinned at the vertices in Sample-I is much stronger than in Sample-III, this increased interaction does not lead to an improved quality of the vortex ice, rather the opposite is observed. In addition to the parameter $P_{ice-pair}$, we introduce the difference $P_{n_{in}=1}-P_{n_{in}=2}$ which also represents the quality of vortex ice. Indeed, this parameter should be zero if the vortex ice is perfect, and different from zero otherwise. Fig. 4(a) shows that the $P_{n_{in}=1}-P_{n_{in}=2}$ ($P_{ice-pair}$) of Sample-I at $H_{1}/2$ is much more (less) than that in Sample-III (Fig. 3c-d). Additionally, $P_{ice-pair}$ is far from 100\% and defects are still observed in the vortex state of Sample-I at $H_{a}^{*}$ ($H_{a}^{*}$ is the magnetic field value where $P_{n_{in}=1}=P_{n_{in}=2}$). Therefore, the quality of the vortex ice in Sample-I is rather poor and this shows that simply reducing the antidot distance is not sufficient.

By comparing Sample-III with the other three samples, the discordant between $L_{1}$ and $L_{2}$ has a significant impact on the increase of $P_{n_{in}=2}$ and decrease of $P_{n_{in}=1}$ as a function of the applied magnetic field in the range of $H_{1}/3 < H_{a} < H_{1}/2$. A direct comparison between samples shows that $P_{n_{in}=1}-P_{n_{in}=2}$ at $H_{1}/2$ is the smallest and $P_{ice-pair}$ is the largest in Sample-III. Moreover, empty-pairs and saturated-pairs are observed simultaneously in the range of $H_{1}/3 < H_{a} < H_{1}/2$ in the samples with $L_{1} > L_{2}$. Therefore, unlike Sample-III, the amount of ice-pairs is not increased and a defect free vortex ice state cannot be realized by simply tuning the magnetic field. This can be explained by the V-V interactions and occupations of vortices. The interaction between vortices at vertex dominates significantly in the samples with $L_{1} > L_{2}$. As such, the occupations of coming vortices are quite easy to result in disorder and defects of empty-pairs and saturated-pairs in such vortex system since such defects lead to relatively small increasement of energy. To conclude, a too large difference between $L_{1}$ and $L_{2}$ is not optimal to observe the vortex ice state.

The center-to-center distance between antidot-pairs $L_{s}$ is the kagome lattice constant, which reflects the averaged interactions between ice-pairs. In the last part, we explore the vortex ice correlations as a function of the kagome lattice constant $L_{s}$. As shown in Fig. 5, the percentage of ice-pairs $P_{ice-pair}$ at $H_{1}/2$ and at $H_{a}^{*}$, $P_{(n_{in}=1)} - P_{(n_{in}=2)}$ at $H_{1}/2$ is not a monotonous function of the kagome lattice spacing $L_{s}$ although the value of $H_{a}^{*}$ decreases monotonously. Therefore, besides the coordination between $L_{1}$ and $L_{2}$, the observed non-monotonic variations of vortex ice also suggest that the optimal kagome lattice for observing vortex ice should not be designed with a too small lattice constant. The experiments by Guillam\'{o}n \emph{et al} \cite{vortex_disorder1} shows that the more disordered vortex states are observed at high magnetic fields due to the quenched disorder. Therefore, the decrease of the quality of vortex ice with reducing the lattice constant is reminiscent of the quenched disorder at high magnetic field.

In summary, direct visualizations of the vortex lattice in superconducting films with a kagome lattice of paired antidots by scanning Hall probe microscopy shows that a vortex ice state starts to develop at $H_{1}/2$ and persists up to $2H_{1}/3$ due to the presence of interstitial vortices. Such unanticipated vortex ice states are more robust against defects than the conventional ice states at $H_{1}/2$. It is found that the vortex ice system is highly tunable by varying magnetic field and the kagome lattice parameters. Beyond the theoretical predictions, a comparison among different designs demonstrates that the defect free vortex ice cannot be attained through enhancing the vortex-vortex interactions via reduction of the size of the kagome lattice. We identified some of the key aspects needed to create a high quality vortex ice systems. Our findings will encourage further theoretical calculation taking into account the presence of interstitial vortices. Moreover, the observed highly tunable vortex ice shows great potential to explore the physics of general ice systems, frustration and order-disorder transitions in complex energy landscapes.

\vspace{2ex}
\noindent
\textbf{Acknowledgments}

C.X. and A.H. acknowledge support by the National Natural Science Foundation of China (Grant No. 11702218 and No. 11702034) and Fundamental Research Funds for the Central Universities (Grants No. G2016KY0305 and No. 310812171011). Y.H.Z., C.X. and A.H. acknowledge the National Natural Science Foundation of China (Grant No. 11421062) and the National Key Project of Magneto-Constrained Fusion Energy Development Program (Grant No. 2013GB110002). J.-Y.G, V.S.Z., J.V.d.V., and V.V.M. acknowledge support from the Methusalem funding by the Flemish government and the Flemish Science Foundation (FWO-Vl). J.-Y.G. also thanks the support by The Program for Professor of Special Appointment (Eastern Scholar) at Shanghai Institutions of Higher Learning. The work of A.V.S. has been supported in part by PDR T.0106.16 of F.R.S.-FNRS. This work is supported by the CA16218 COST Action.

\vspace{2ex}
\noindent
xuecun@nwpu.edu.cn

\noindent
Junyi$\_$Ge@t.shu.edu.cn

\noindent
joris.vandevondel@kuleuven.be

\end{document}